# The perceived quality of process discovery tools

Francis Bru and Jan Claes

Department of Management Information Science and Operations Management,
Faculty of Economics and Business Administration,
Ghent University, Belgium
{francis.bru,jan.claes}@ugent.be

**Abstract.** Process discovery has seen a rise in popularity in the last decade for both researchers and businesses. Recent developments mainly focused on the power and the functionalities of the discovery algorithm. While continuous improvement of these functional aspects is very important, non-functional aspects such as visualization and usability are often overlooked. However, these aspects are considered valuable for end-users and play an important part in the experience of these end-users when working with a process discovery tool. A questionnaire has been sent out to give end-users the opportunity to voice their opinion on available process discovery tools and about the state of process discovery as a domain in general. The results of 66 respondents are presented and compared with the answers of 63 respondents that were contacted through one particular software vendor's employee and customer base (i.e., *Celonis*).

## 1      Introduction

Since the release of the Process Mining Manifesto six years ago [1], the world of process mining has seen a massive increase in end-users in both the research and the business domain. This increase of attention can be observed in facts such as extending the International Workshop on Business Process Intelligence to a two day event since 2017 [2]. The majority of the research presented in the workshop is centered around the creation of more accurate or faster algorithms, the development of new functionalities, or describes new case studies in terms of effectiveness and efficiency of various process mining projects. The functional aspects of process mining that are targeted by these works are important for fine-tuning this fairly new technology.

Unfortunately, this appears to mean that non-functional aspects of process mining tools, such as visualization and user-friendliness, are placed in the background. Interestingly, software quality models include these non-functional aspects as key metrics for the quality of software [3]. Moreover, an earlier study revealed that end-users of process mining find these non-functional aspects equally important as the functional aspects [4]. To shed light on the preferences of the end-users, a questionnaire was created in which we collected perceptions from the community about aspects of current process discovery tools.

This paper complements studies on process mining quality metrics and models for software quality. In 2011, Van der Aalst described fitness, generalization, precision, and simplicity as key quality criteria for process mining results [5]. Other studies, have further described and developed subcategories for these criteria [6, 7] and have developed a benchmark to assess the quality of the discovered process model [8]. Whereas in the process mining field, the non-functional aspects of the algorithm have not gained much attention, in the field of manual process modeling, the focus appears to be exactly on these aspects (see a systematic literature review of studies on business process modeling quality [9]). On the contrary, researchers have only recently begun to explore the functional aspects of modeling in a research stream called the process of process modeling [10, 11]. However, these studies focus only on the discovered model and not on the entire process discovery tool.

When evaluating process mining tools and techniques from a software quality point of view, non-functional aspects appear to be prioritized over functional aspects [3, 12]. One particular illustrative example is the ISO/IEC 25010 standard, which includes quality dimensions such as UI aesthetics, understandability, changeability, etc. [13]. Also two short market studies have been published, one in 2017 by Devi, et al., comparing the functionalities of *ProM*, *Disco* and *Celonis* [14], and one very recently published by *Gartner* in which the process mining market is analyzed and the market approach of 15 tools are listed [15].

This paper is structured as follows: Section 2 describes the applied research methodology. Section 3 presents the main results, which are further discussed in Section 4.

## 2      Methodology

The research goal is to gain insight in how end-users of process mining experience the different aspects of process discovery tools. The focus is put on process discovery and not on process mining in general because it is considered the most typical [16] and most adopted [15] technique of process mining. Additionally, most commercially available tools tend to concentrate on process discovery and less on conformance checking or enhancement [17]. Data about the opinion of the end-users was collected through an online questionnaire.

The first four questions map the demographical background of the respondents (i.e., age, occupation, geographical location, and experience with process discovery). After these demographical questions, the respondents were presented a grid in which their familiarity with the different identified process discovery tools was questioned through Likert scales. The questions that followed gave the respondents a chance to criticize or to give appreciation to the tools they knew. Their opinion was measured in two different ways: a 5-star scoring system on one functional and three non-functional categories (i.e., analytical performance, user friendliness, visualization, and safety & security), and two open-ended questions in which they could elaborate on the benefits and on the drawbacks of the tool respectively. Finally, the respondents had to answer 3 short questions: two about the offline or online environment of the tool (preference and reason of their choice) and one about the state of process discovery in general by positioning it in the technology adoption lifecycle.

The questionnaire was released on December 8, 2018 and was made available for the next 80 days. A call for participation was posted in the process mining groups of LinkedIn and Facebook. Respondents were also contacted personally through email and LinkedIn message, and suggested to share the survey with colleagues and friends with interest and experience in process discovery. The personal approach proved to be fruitful in convincing respondents to participate in the study. This resulted in 66 respondents completing the entire survey. The question in the survey with a maximum of answers received 89 answers and the minimum was 30 answers (an optional question, which was skipped by the majority of respondents). This results in an average of 70 answers per question.

A copy of the questionnaire can be consulted at janclaes.info/papers/PDSurvey.

## 3   Results

In this section, a selection of the most interesting results of the survey are listed and discussed. The raw data can be found at janclaes.info/papers/PDSurvey.

### 3.1   Demographics

The demographical data shows that a heterogeneous group of respondents is formed:

- Process discovery is used in a variety of different professions. Over 75 percent of the respondents use process discovery for a commercial purpose. The other respondents use process discovery in their studies or for a research subject (cf. Fig. 1a).

- The respondents' age varies between 22 and 64 years. The majority of respondents is between 22 and 40 years old (cf. Fig. 1b).

- The respondents are situated all over the world: 26 countries are represented in this survey. There is a high concentration of respondents in Belgium and the Netherlands (cf. Fig. 1c). Two explanations can be given for this clustering. (1) Both authors are located in Belgium. When contacting respondents personally to join the survey, a message in Dutch was used when the respondent indicated on their LinkedIn profile to speak this language. Being contacted in their native tongue may have convinced people to join the survey. (2) The Netherlands is the leader in research on this subject, resulting in a higher concentration of researchers and users in this area.

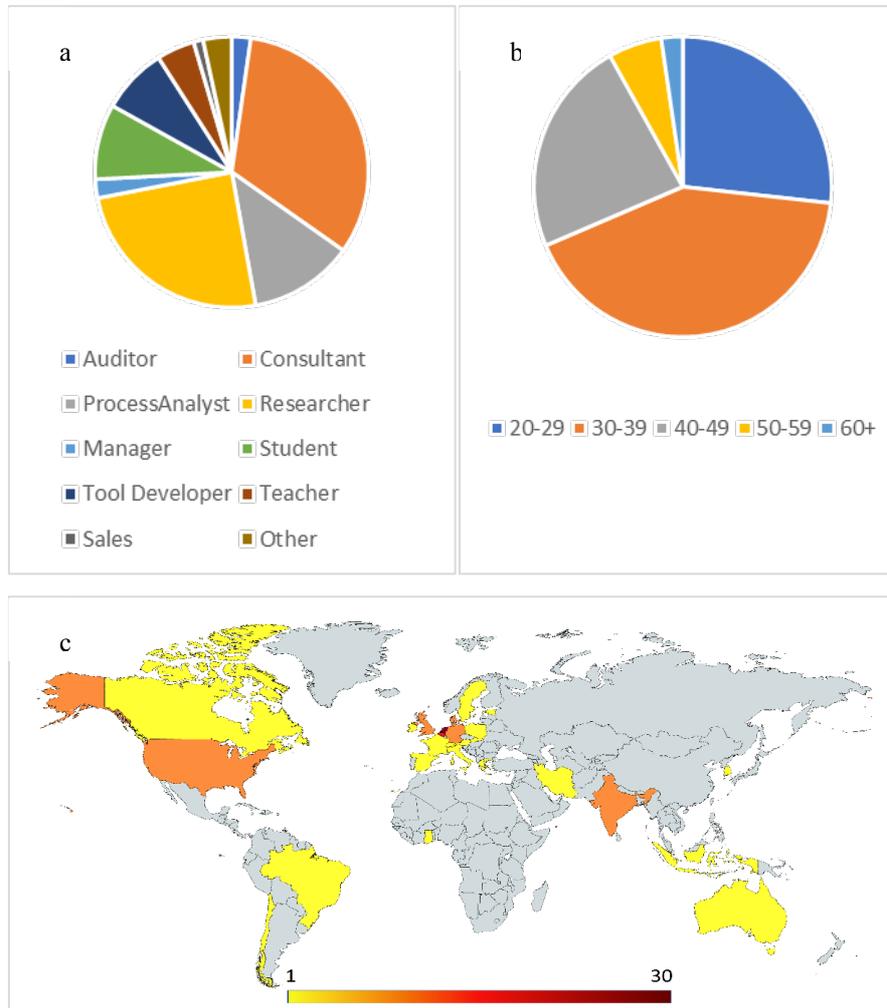

Fig. 1. Demographics of the respondents
(question 1 (86), 2 (87) and 3 (89 respondents), map created at mapchart.net)

### 3.2 Popularity of different process discovery tools

The most popular tools for process discovery are *Disco* and *ProM* (cf. Fig. 2). It is clear that a lot of new developers have joined the market over the recent years. Seven of the selected tools were not yet described in a process mining manual from 2011 [5] and five were not mentioned in a similar survey research from 2014 [4]. The latter reference reports also that *Disco* and *ProM* are the market leaders, however *Celonis Process Mining* has risen from a rather unknown tool to a third-place spot on this list. This top 3 applies to every part of the world according to the collected data.

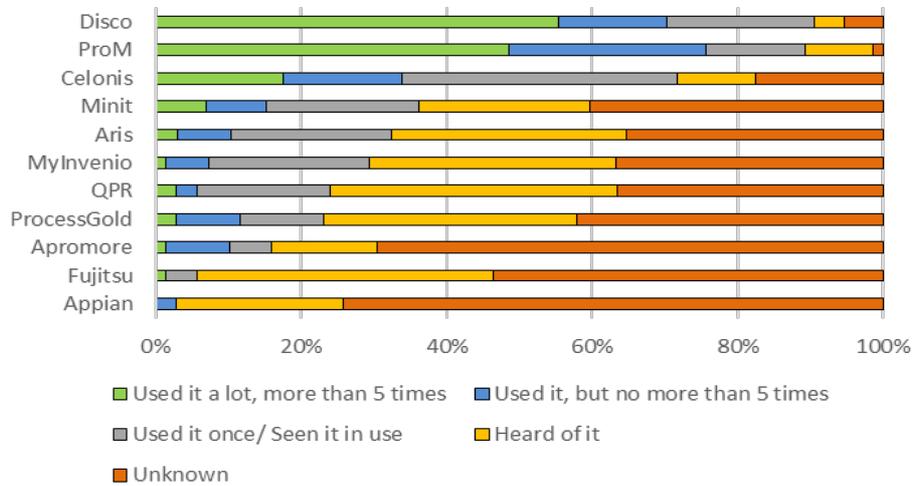

Fig. 2. Popularity of process discovery tools (question 5, 74 respondents)

### 3.3 Importance of tool aspects

When the respondents were asked what they like or dislike about the tools they were familiar with, intuitiveness and ease to use (or the opposites: complex and not user friendly) is mentioned the most (cf. Fig. 3). The second most listed aspect is functional: the number (or lack) of features. These are followed by process map visualization and integration with other analysis tools. Concerning the latter aspect, respondents mentioned they prefer the tool to be integrated in a suite in which other BPM tools can be used as well. These four aspects are used to compare the most popular tools according to the previous segment.

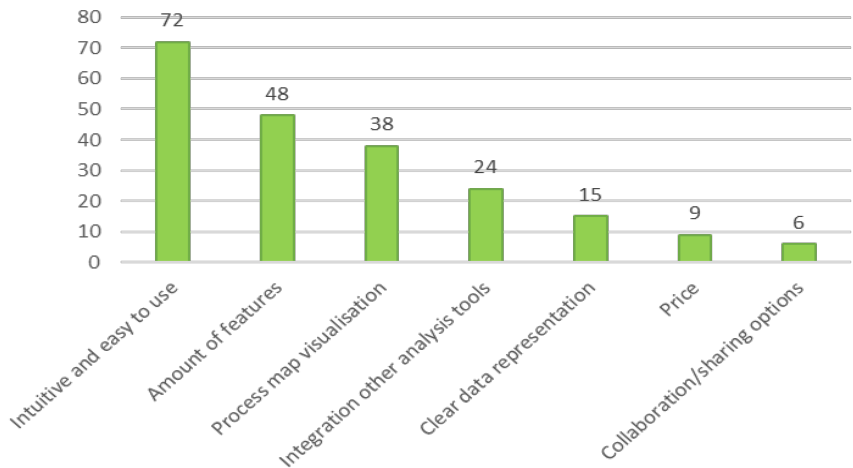

Fig. 3. Most important aspects of the process discovery software
(question 6 (60) and question 7 (40 respondents))

### 3.4 Differences between the most popular tools

When comparing the two most popular tools (cf. Section 3.1), we observe a difference that is probably related to their market strategy. *ProM* an open-source platform for researchers and more advanced users with a plethora of user created functionalities. In contrast, *Disco* focusses on novice users and provides a basic but easily understandable tool. This distinction is also reported by the respondents, giving positive comments on *Disco*'s intuitiveness and ease to use and indicating the lack of features as their biggest drawback (cf. Fig. 4 and 5).

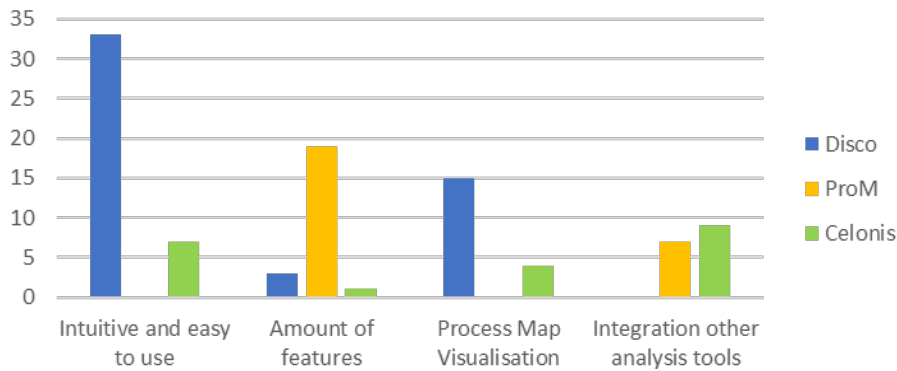

Fig. 4. Benefits of the three most popular process discovery tools
(question 6 (60) and question 7 (40 respondents))

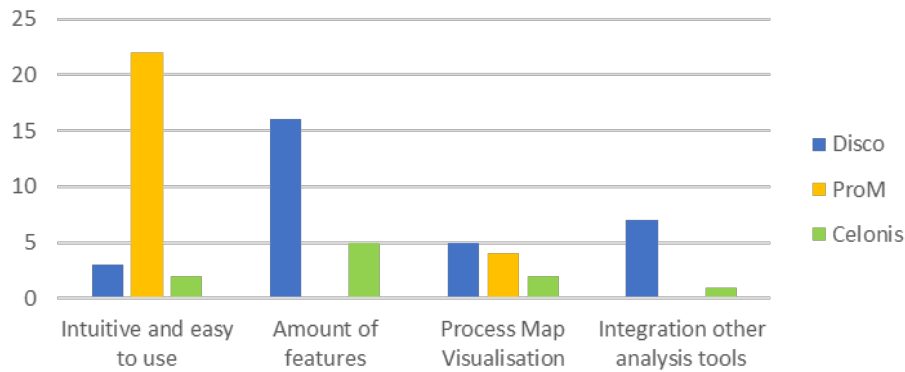

Fig. 5. Drawbacks of the three most popular process discovery tools
(question 6 (60) and question 7 (40 respondents))

The reverse can be said for *ProM*. It is praised for its many features, but it is perceived as complicated and not user friendly. This is confirmed by another research survey on the *ProM* framework [18]. Respondents' opinions appear to follow a reverse relation between the complexity of the tool and the number of features it has, with *ProM* and *Disco* being the two extremes of the spectrum. Respondents expressed the lack of a user-friendly and good-looking tool that helps them to advance their expertise in the field of process discovery.

Other process discovery tools may fill in this gap. *Celonis Process Mining* for example, receives both negative and positive comments on all these aspects that were indicated to be important. This means that they fall somewhere in between *ProM* and *Disco* in this spectrum. Indeed, *Celonis* focusses more on process discovery at a business level. They were the only tool to receive positive comments on their collaboration and sharing options.

### 3.5 Using an online or offline interface

The tool developers show different approaches for hosting their tool. *ProM* is available through software that has to be installed on a computer, *Celonis* chooses a web client for hosting their tool and *Disco* provides both options. A third identified approach is an application for Excel, such as the *QPR ProcessAnalyzer* tool. When our respondents were asked which of these approaches they prefer, the web client and the standalone program are almost tied (cf. Fig. 6). The Excel application is only preferred by 4 of the 70 respondents.

When asking for the cause of their preference, respondents proclaimed processing speed as an important aspect for both the offline and the online interface. Respondents also indicate that a standalone program may be safer and more secure because the data is stored locally and not in the cloud. Furthermore, we see the obvious reason for this preference: no installing and access on any computer for the online environment and not needing internet to access the tool for the offline environments. The integration or combination of the process discovery tool with other BPM tools is once again an aspect the respondents indicate to be important (cf. Section 3.3).

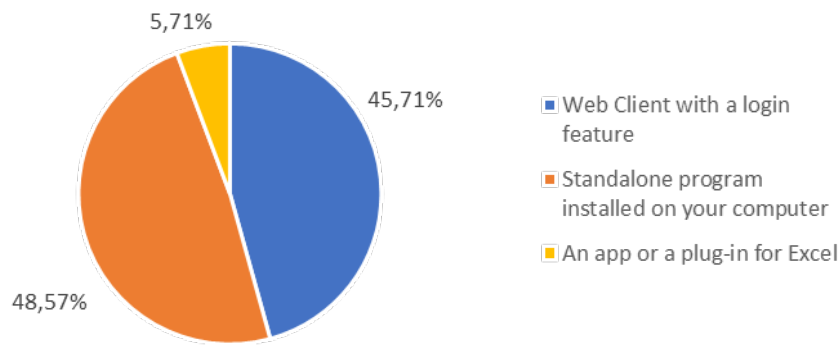

Fig. 6. Software environment preference (question 9, part 1, 70 respondents)

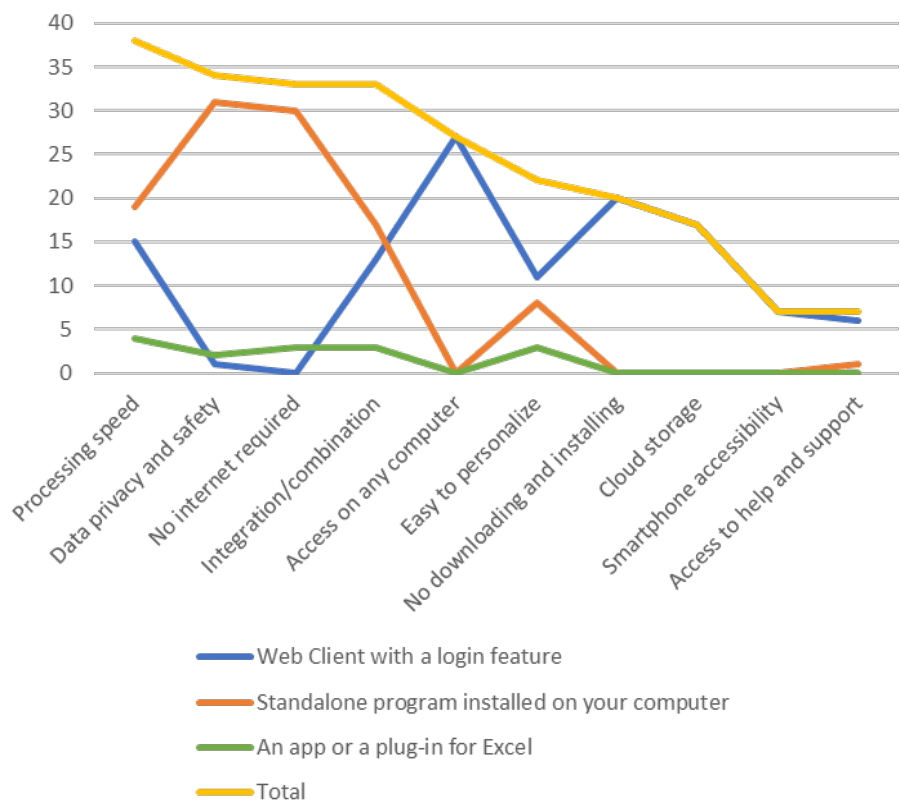

Fig. 7. Benefits of the software environment (question 9, part 2, 70 respondents)

### 3.6 State of process discovery in general

In a final question, the respondents were asked to indicate in which phase of the technology adoption lifecycle they would place process discovery at the moment. More than half of the respondents classified process discovery in the "early adopters" phase, followed by the "early majority" phase as second biggest group (about 27 %). Meade & Rabelo [19] describe a "chasm" between these two phases, because a high number of products fail at this stage. They suggest that a focus on the end-user and providing in the needs of a certain niche market are the best strategies to ensure the continuity of the product. A recently published market study reveals that each vendor has his own strongpoints to gain a competitive advantage in the process mining market [15]. This strategy is necessary to attract new users in this growing market [20].

### 3.7 Opinion of a tool developer

When distributing the survey, software developers were contacted to gain insights in how they try to impact the process discovery market. *Celonis* was really interested in a cooperation and shared our survey through its internal mailing list. This resulted in a second data set consisting of 63 respondents of which about 90 % are employees and 10 % clients of *Celonis*. The answers of this additional group of respondents was used to determine if the software developer's market strategy aligns with the wishes of the end-users and the market strategy suggested by the technology adoption lifecycle (cf. Section 3.6).

When comparing the positioning of process discovery on the technology adoption lifecycle, both groups agree on the "early adopters" and "early majority phase (cf. Fig. 8). In other words, Celonis should realise that a focus on end-users and a niche market strategy are needed for success.

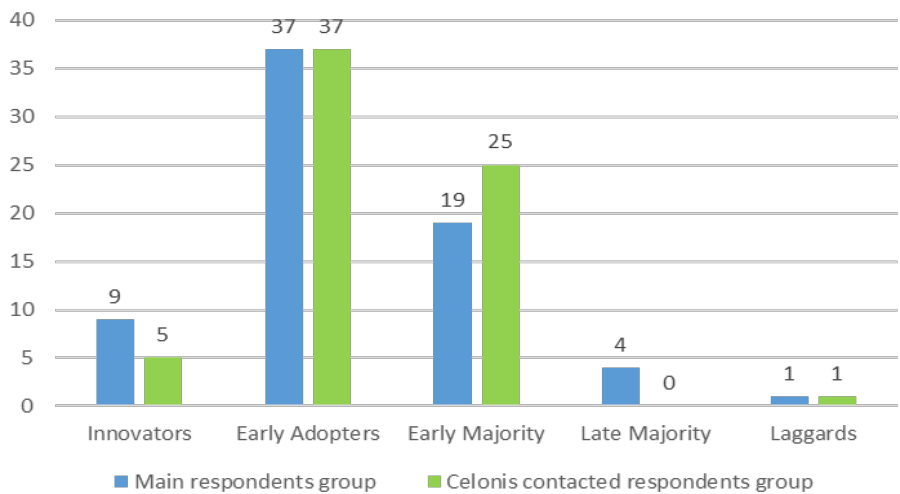

Fig. 8. Positioning of process discovery in the technology adoption lifecycle
(question 10, main group: 70 respondents, *Celonis* contacted group: 68 respondents)

The respondents group contacted by *Celonis* prioritize user friendliness and the number of features (cf. Fig. 9), just like the respondents from the general survey (cf. Section 3.3). Additionally, two different aspects are mentioned by the *Celonis* specific group: Enterprise focus and the ability to analyze large datasets. *Celonis* appears to use these two aspects as differentiator to gain a competitive advantage on the process mining tool market. However, when looking at the *Celonis* website, these aspects are not mentioned as one of their strengths. The main advertising strategy for their product is showing advantages of process mining in general (i.e. visualizing process, identifying faults and risks, etc.). According to the technology adoption lifecycle, this will not suffice in the growing market in which convincing customers of your niche strengths will be needed to survive (cf. Section 3.6).

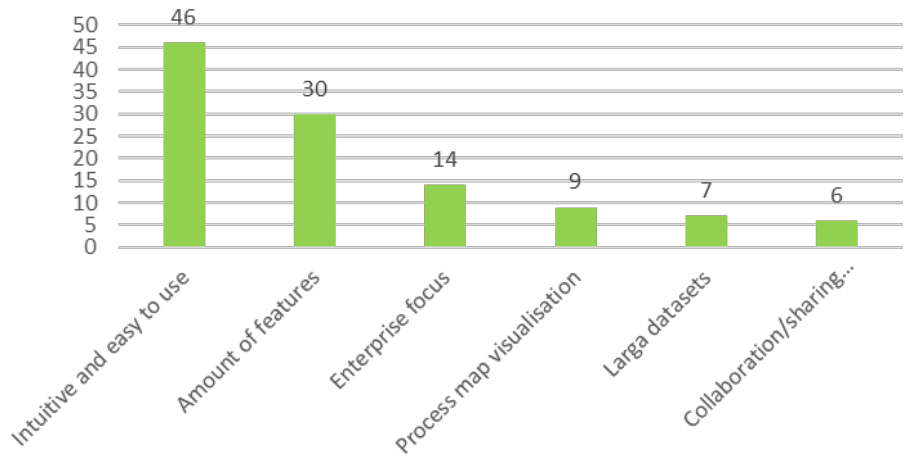

Fig. 9. Most important aspects of the *Celonis* Process Mining tool as perceived by the respondent group contacted by *Celonis* (58 respondents)

Finally, concerning the software environment, nearly all of *Celonis'* respondents prefer a web client (63 out of 69 answers). The main causes for this preference for an online tool are similar to those of the general respondents group (cf. Section 3.5) with "access on any computer", "processing speed", "no downloading and installing" and "cloud storage" being the four most important aspects.

## 4 Conclusion

Using an exploratory survey, the preferences of end-users and their perception of the quality of process discovery tools were analyzed. This resulted in the suggestion of focus points and market strategies for the software developers trying to gain a competitive advantage.

Firstly, we determined that *Disco* and *ProM* are still the most well-known vendors. However, new competitors have joined in the recent years, finding more and more popularity in this growing market.

The next sections outlined which aspects of process discovery tools are key factors for the end-users. We observe that user friendliness, the number of features, visualization and the integration with other analysis tools are considered the most important aspects.

When comparing the drawbacks and benefits of the most popular tools (i.e., *Disco* and *ProM*) a reverse relation between the complexity of the tool and the number of features is laid bare, with *Disco* and *ProM* being on different sides of this spectrum.

Considering the different possible approaches for the software environment, both the offline and online options are preferred by nearly equal amounts of respondents. The main causes for their preference is safety and security and not needing internet for the offline options and no installing and access on any computer for the online environment. Processing speed is an important aspect for both options.

Respondents were also asked to position process discovery on the technology adoption lifecycle in which they indicated "early adopters" and "early majority" as the main classifications. This resulted in a market strategy suggestion: focus on the end-user and provide in the needs of a certain niche market.

In a final part, the cooperation of software vendor *Celonis* allowed us to compare the insights of the paper with the strategy and focus of a developer. Here we conclude that the aspects, they focus primarily, are similar to the respondents' preferred aspects, along with some additional differentiating aspects, giving them the possibility of a niche approach. However, looking at the *Celonis* website, the general process mining advantages are still used in promoting their product.

This research paper shows that non-functional aspects of process discovery tools can play an important part in the end-users decision-making process when choosing a tool. Vendors and developers of these tools can find some useful insights such as focus points or how to position themselves in the market. In this case, the proposed market strategy for process discovery tools is niche marketing. Vendors may find a competitive advantage by marketing their tool to a specific segment of end-users. Process discovery users are interested in changes in this technology (over 80 % of the respondents subscribed to the mailing-list for this paper) so they are likely to notice a change of strategy. Additionally, new users are advised to compare different vendors when choosing their BPM solution [21], so vendors having a unique selling point are more likely to gain more attention.